\documentclass[12pt]{article}
\pdfoutput=1
\usepackage[square,comma,numbers,sort&compress]{natbib}
\usepackage{graphicx,epstopdf,amssymb,amsfonts,amsmath,amsthm,array,
mathrsfs,amscd}
\usepackage[vcentermath]{youngtab}
\usepackage{float}
\newcommand{\pbs}[1]{\let\temp=\\#1\let\\=\temp}
\numberwithin{equation}{section}
\def\be{\begin{equation}}\def\ee{\end{equation}}
\def\cvp{\raise 2pt\hbox{,}} 
\def\div{\mathop{\text{div}}\nolimits}

 \def\d{{\rm d}}\def\nn{{\cal
N}}

\def\gs{g_{\text s}}\def\ls{\ell_{\text s}}

\def\Sb{S_{\text b}}\def\Sg{S_{\text g}}

\def\rh{r_{\text h}}
\def\SDBI{S_{\text{DBI}}}\def\SCS{S_{\text{CS}}}\def\SCT{S_{\text{CT}}}
\def\plb#1#2#3{{\it Phys.\ Lett.\ }{\bf B #1} (#2) #3}
\def\npb#1#2#3{{\it Nucl.\ Phys.\ }{\bf B #1} (#2) #3}

\def\jhep#1#2#3{{\it J. High Energy Phys.\ }{\bf #1} (#2) #3}
\def\prd#1#2#3{{\it Phys.\ Rev.\ }{\bf D #1} (#2) #3}

\def\atmp#1#2#3{{\it Adv.\ Theor.\ Math.\ Phys.\ }{\bf #1} (#2) #3}
\def\cmp#1#2#3{{\it Comm.\ Math.\ Phys.\ }{\bf #1} (#2) #3}

\def\mpla#1#2#3{{\it Mod.\ Phys.\ Lett.\ }{\bf A #1} (#2) #3}

\def\cqg#1#2#3{{\it Class.\ Quant.\ Grav. }{\bf #1} (#2) #3}

\def\imath#1#2#3{{\it Invent math }{\bf #1} (#2) #3}

\def\cag#1#2#3{{\it Comm.\ Anal.\ Geom.\ }{\bf #1} (#2) #3}
\def\amath#1#2#3{{\it Adv.\ Math.\ }{\bf #1} (#2) #3}

\begin{document}
%
%
{\pagestyle{empty}
\begin{flushright} MIT-CTP/4613 \end{flushright}
\parskip 0in
\

\vfill
\begin{center}
%
%

{\LARGE Holography, Probe Branes}

\bigskip

{\LARGE and Isoperimetric Inequalities}

\vspace{0.4in}

Frank F{\scshape errari}$^{*}$ and Antonin R{\scshape ovai}$^{\dagger}$
\\

\medskip

{$^{*}$\it Service de Physique Th\'eorique et Math\'ematique\\
Universit\'e Libre de Bruxelles and International Solvay Institutes\\
Campus de la Plaine, CP 231, B-1050 Bruxelles, Belgique}

\smallskip

{$^{\dagger}$\it Arnold Sommerfeld Center for Theoretical Physics\\
Ludwig-Maximilians-Universit\"at M\"unchen\\
Theresienstrasse 37, D-80333 M\"unchen, Deutschland\\
and\\
Center for Theoretical Physics\\
Massachusetts Institute of Technology\\
Cambridge, MA 02139, USA
}

\smallskip
{\tt frank.ferrari@ulb.ac.be, arovai@mit.edu}
\end{center}
\vfill\noindent

In many instances of holographic correspondences between a $d$ dimensional boundary theory and a $d+1$ dimensional bulk, a direct argument in the boundary theory implies that there must exist a simple and precise relation between the Euclidean on-shell action of a $(d-1)$-brane probing the bulk geometry and the Euclidean gravitational bulk action. This relation is crucial for the consistency of holography, yet it is non-trivial from the bulk perspective. In particular, we show that it relies on a nice isoperimetric inequality that must be satisfied in a large class of Poincar\'e-Einstein spaces. Remarkably, this inequality follows from theorems by Lee and Wang.

\vfill

\medskip
%
\begin{flushleft}
\today
\end{flushleft}
%
\newpage\pagestyle{plain}
\baselineskip 16pt
\setcounter{footnote}{0}

}


\newpage
\section{\label{IntroSec} Introduction}

Consider a holographic correspondence between a $(d+1)$-dimensional bulk gravitational theory on a conformally compact manifold $M$ and a $d$-dimensional field theory on its compact boundary $X=\partial M$.\footnote{Physically relevant non-compact manifolds can usually be obtained by taking the large volume limit of a compact manifold.} Assume that the correspondence follows by considering the near horizon limit of a large number $N$ of BPS $(d-1)$-branes (mentioned simply as branes in the following) \cite{malda}. The boundary field theory has $N$ colors and can be interpreted as living on these branes. It is then natural to study the physics associated with probe branes in the bulk geometry. These branes are clearly special since, in some sense, they make up the bulk holographic space itself. Such studies have appeared many times in the literature; particularly instructive results were discussed, for example, in \cite{KT}. 

Recently, a precise construction of the probe brane action $\Sb$ from the point of view of the boundary field theory was proposed \cite{fer1}. The main motivation in \cite{fer1} is to provide purely field theoretic tools to study holography in a wide range of models. It is shown that the probe action naturally describes the motion of the brane in a higher dimensional holographic bulk space, including in the case of the pure Yang-Mills theory where a fifth dimension automatically emerges \cite{fer1}. In particular, the details of the bulk geometry can be read off from the probe action \cite{fer2}. 

The construction in \cite{fer1} implies an interpretation of the probe brane that seems to depart from the standard lore, which relates the presence of a probe brane in the bulk to some Higgsing of the gauge group on the boundary. Instead, the Euclidean partition function for $K$ probe branes in the bulk is shown in \cite{fer1} to compute exactly the ratio $Z_{N+K}/Z_{N}$ between the Euclidean partition functions of the boundary theory for $N+K$ and $N$ colors respectively,
\be\label{pathint} \frac{Z_{N+K}}{Z_{N}} = \int\!\mathscr D\Sigma\, e^{-\Sb(\Sigma)}\, ,\ee
where we have denoted by $\Sigma$ the degrees of freedom living on the brane. This point of view has many interesting consequences and seems consistent with the notion of Highly Effective Action described in \cite{HEA}, which corresponds to the special case $N=K=1$.

The aim of the present work is to understand, from the bulk perspective, one of the simplest consequence of Eq.\ \eqref{pathint}. Assume that the free energy $-\ln Z_{N}$ scales as $N^{\gamma}F$ at large $N$, for some exponent $\gamma$, with corrections $o(N^{\gamma -1})$ (for example, in the standard gauge theories considered in \cite{fer1}, $\gamma=2$ and the corrections are of order $O(N^{0})=o(N))$. Then $\ln(Z_{N+1}/Z_{N})=-\gamma N^{\gamma-1} (F + o(1))$. On the other hand, in the large $N$ limit, the probe brane action $\Sb$ is very large (for example, it is proportional to $N$ in gauge theory). The right-hand side of \eqref{pathint} is then dominated by configurations minimizing $\Sb$. If we denote by $\Sb^{*}$ the minimum value of $\Sb$, we obtain in this way $\ln(Z_{N+1}/Z_{N}) = -\Sb^{*}$. If, moreover, we use the standard holographic dictionary of \cite{shol1,shol2} that identifies $N^{\gamma}F$ with the on-shell gravitational bulk action $\Sg^{*}$, we get the fundamental identity
\be\label{fundid} \Sg^{*} = \frac{N}{\gamma}\Sb^{*}\, .\ee
This is an archetypal holographic identity, relating a bulk quantity on the left-hand side to a surface quantity on the right-hand side.

The reasoning that leads to \eqref{fundid} is very robust and we believe that it constitutes an important basic property of holography. However, it does not constitute a proof. Indeed, it assumes that the gauge theoretic probe brane action constructed in \cite{fer1} (for which the identity \eqref{fundid} is a rigorous mathematical statement) matches with the standard bulk notion of probe brane action. This is a new addition to the gauge theory/string theory dictionary and, as any other entry in this dictionary, it cannot be rigorously proved. Our aim in the following will be to test this proposal, by deriving \eqref{fundid} \emph{directly from the bulk perspective,} thus without using the results of \cite{fer1} or assuming the validity of \eqref{pathint}.

Let us note that the relation \eqref{fundid} can be easily generalized to a large variety of situations, including cases with several types of branes (like for example in $\text{AdS}_{3}$ holography), cases where $\alpha'$ corrections and finite $N$ effects are included, and even cases corresponding to asymptotically flat geometries. We let the discussion of some of these extensions to a companion paper \cite{FR} and focus presently on the basic conceptual issues in the simplest framework.

At first sight, the equality \eqref{fundid} seems rather puzzling, for at least two basic reasons. First, the gravitational action is naively infinite and a holographic renormalization procedure is required to make sense of it \cite{holoren}, whereas the probe brane action is naively finite with no need to renormalize. Yet, equation \eqref{fundid} implies that an analogue of the holographic renormalization prescription must exist for the probe brane action and we have to understand what this could be. Second, the gravitational action $\Sg$ is the sum of the bulk Einstein-Hilbert term and a boundary Gibbons-Hawking term, whereas the brane action $\Sb$ is a purely surface quantity, sum of Dirac-Born-Infeld and Chern-Simons contributions.  To compute the on-shell value $\Sb^{*}$, one naively has to solve the field equations on the brane and evaluate $\Sb$ on the solution. This looks quite complicated and the matching with the very different-looking on-shell gravitational action may seem rather miraculous. Clearly, in view of the claimed extreme generality of \eqref{fundid}, a simple mechanism must be at work, simplifying drastically the analysis and ensuring consistency.

We shall elucidate these issues in the following in the case of pure gravity, where the bulk space $M$ is a Poincar\'e-Einstein manifold,
\be\label{PEcond} R_{\mu\nu} = -\frac{d}{L^{2}} G_{\mu\nu}\, .\ee
We shall prove that the consistency of \eqref{fundid} relies on a non-trivial isoperimetric inequality, bounding from below the area $A(\Sigma)$ of any hypersurface $\Sigma\subset M$ homologous to the boundary by the volume $V(M_{\Sigma})$ of bulk space enclosed by $\Sigma$,
\be\label{isop} A(\Sigma) \geq \frac{d}{L} V(M_{\Sigma})\, .\ee
It is easy to see that this inequality is violated if the Yamabe constant of the boundary is negative. Holography thus cannot be consistent in these cases, a fact that has been known for a long time \cite{WY,Yampos} (a negative Yamabe constant simply means that the action for a conformally coupled scalar on the boundary will not be bounded from below, implying that the field theory on the boundary is ill-defined). Quite remarkably, when the Yamabe constant is non-negative, the inequality \eqref{isop} can be derived from the details of the proof of a theorem by Lee \cite{Lee} and was also proved directly by Wang in \cite{Wang}.

\section{\label{ExSec} A simple example: Schwarzschild-$\text{AdS}_{5}$}

It is very useful to first analyse a simple example. So let us consider the famous Schwarzschild black hole in $\text{AdS}_{5}$, which is dual to the $\nn=4$ gauge theory on $X=\text{S}^{3}\times\text{S}^{1}$, when the temperature is above the Hawking-Page transition \cite{shol2,WSch}. We pick the standard representative
\be\label{barg1} \bar g = \d t^{2} + a^{2}\d\Omega_{3}^{2}\ee
for the conformal class of the metric on $X$, where $t$ and $t+\beta$ are identified and $\d\Omega_{3}^{2}$ is the round metric of radius one on $\text{S}^{3}$. The bulk metric can be conveniently written by using the Fefferman-Graham coordinates associated with \eqref{barg1} as\footnote{The use of Fefferman-Graham coordinates near the boundary will make the general discussion in the next section easier. In the present case, these coordinates cover the full bulk manifold.}
\be\label{AdSSchmet} G = \frac{1}{r^{2}}\Bigl[L^{2}\d r^{2} + f(r)^{-1}\bigl(1-(r/\rh)^{4}\bigr)^{2}\d t^{2}+ a^{2}f(r)\d\Omega_{3}^{2}\Bigr]\, \cvp\ee
with
\be\label{fSAdef} f = 1-2\alpha x + x^{2}\, ,\quad x = (r/\rh)^{2}\, ,\quad \alpha = \frac{L^{2}\rh^{2}}{4 a^{2}}\,\cdotp\ee
The full cigar-shaped bulk manifold $M=\text{B}^{2}\times\text{S}^{3}$ is covered when $0<r\leq\rh$ (or $0<x\leq 1$), with $r=0$ corresponding to the boundary and $r=\rh$ to the tip of the cigar (horizon). The parameter $\alpha$ belongs to the interval $]0,1[$, ensuring that $f>0$. Smoothness at $r=\rh$ yields the relation
\be\label{betaalpharel} \beta=\pi a\sqrt{2\alpha (1-\alpha)}\ee
between $\alpha$ and the inverse temperature $\beta$.

Let us now consider a 3-brane, which is a hypersurface $\Sigma$ in $M$. In the present section, for simplicity and consistently with the symmetries of the metric \eqref{AdSSchmet}, we limit our discussion to hypersurfaces given by an equation $r=\text{constant}$. The brane action is then a function of $r$, sum of DBI and CS contributions. The DBI term is simply the area of the hypersurface for the induced metric times the 3-brane tension. A simple calculation yields
\be\label{SDBIex} \SDBI(r) =\frac{2\pi^{2}\tau_{3}a^{3}\beta}{\rh^{4}}\,\frac{(1-x^{2})(1-2\alpha x + x^{2})}{x^{2}}\,\cdotp \ee
This term is a monotonically decreasing function of $x$ (or of $r$). It tends to make the brane shrinks. The Euclidean CS term is
\be\label{SCSdef} \SCS = -i\tau_{3}\int_{\Sigma}\! C_{4}\, ,\ee
where the Ramond-Ramond five-form field strength $F_{5}=\d C_{4}$ is related to the bulk volume form $\Omega_{5}$ by
\be\label{RR5form} F_{5} = \frac{4i}{L}\Omega_{5}+\cdots\ee
The $\cdots$ represent components on the $\text{S}^{5}$ part of the ten-dimensional geometry, which must be present because $F_{5}$ is self-dual. However, these terms play no role in our discussion, nor does the $\text{S}^{5}$. This is why we have not mentioned them up to now, and we shall not mention them any longer. It is straightforward to integrate the volume form of the metric \eqref{AdSSchmet} to obtain $C_{4}$. \emph{The integration generates and arbitrary integration constant $c$,} yielding
\be\label{C4AdS} C_{4} = ia^{3}\Bigl(-\frac{1}{r^{4}}+\frac{L^{2}}{a^{2}r^{2}}+\frac{L^{2}r^{2}}{a^{2}\rh^{4}}-\frac{r^{4}}{\rh^{8}}+c\Bigr)\d t\wedge \omega_{3}\, ,\ee
where $\omega_{3}$ is the volume form on the unit radius round 3-sphere. Plugging into \eqref{SCSdef}, we get
\be\label{SCSex} \SCS(r) = -\frac{2\pi^{2}\tau_{3}a^{3}\beta}{\rh^{4}}\,\frac{x^{4}-4\alpha x^{3}-4\alpha x + 1}{x^{2}} + s\, ,\ee
for some $x$-independent constant $s$ (which is proportional to the constant $c$ in \eqref{C4AdS}). The CS term is a monotonically increasing function of $x$, tending to make the brane inflate towards the boundary $x=0$. Adding up \eqref{SDBIex} and \eqref{SCSex}, we finally get
\be\label{Sbex} \Sb(r) = \frac{4\pi^{2}\tau_{3}a^{3}\beta}{\rh^{4}}\,\frac{\alpha + 3\alpha x^{2}-x^{3}}{x} + s\, .\ee

This formula has three important basic qualitative features. First, it is a monotonically decreasing function of $x$: the DBI term wins over the CS term and the brane wants to shrink. The minimum value of the action is obtained for the maximum value $x=1$ of the variable $x$, for which the shrunken brane sits at the tip of the cigar,
\be\label{Sbexstar} \Sb^{*} = \frac{4\pi^{2}\tau_{3}a^{3}\beta}{\rh^{4}}\bigl(4\alpha -1 \bigr) + s\, .\ee
Second, $\d\Sb/\d r <0$ at $x=0$: the brane equations of motion are not satisfied at the minimum of the action, and, actually, have no solution! Third, \emph{the result depends on an arbitrary constant $s$.}

For most purposes, this ambiguous constant $s$ in the brane action is inoffensive. It can be interpreted as coming from the gauge symmetry $C_{4}\mapsto C_{4}+c_{4}$, for any closed 4-form $c_{4}$. However, for our purposes, it clearly does play a crucial role. Our aim is to find the on-shell value $\Sb^{*}$ of the brane action and any undetermined constant would allow to shift $\Sb^{*}$ to any value we like, which is of course nonsense. 

A na\"ive way to fix the constant $s$ in \eqref{Sbexstar}, or, equivalently, the constant $c$ in \eqref{C4AdS}, could be to impose the global regularity of the potential $C_{4}$. This would imply that the term in parenthesis in \eqref{C4AdS} has to vanish when $r=r_{\text h}$. However, this condition is artificial and, as we shall see, utterly incorrect. There is no reason to impose a global regularity condition on a non gauge-invariant object. Only the field strength $\d C_{4}$ must be globally defined, and of course it is for any choice of the constant $c$. 

This situation is, actually, quite familiar, at least in the context of asymptotically flat black hole solutions. For example, the gauge potential for a Euclidean four-dimensional Reissner-Nordstr\"om black hole of charge $Q$ reads $A=-i(Q/r +c)\d t$ in standard coordinates. The constant $c$ is indeterminate, but is most naturally chosen such that the electrostatic potential of a charged probe particle vanishes at infinity. This yield $c=0$. This is a very natural and physically sound condition, simply stating that the energy of a particle in flat space should be given only by its rest mass with no constant contribution from an electrostatic potential at infinity. It implies that the gauge potential $A$ is \emph{not} globally defined, since this would inconsistently imply $c=-Q/r_{+}\not = 0$, where $r=r_{+}$ is the horizon.

What is happening in our asymptotically AdS set-up is actually very similar.\footnote{This similarity can be made extremely precise in some cases, when the asymptotically AdS geometry is obtained from the near-horizon limit of an asymptotically flat geometry, see \cite{FR}.} The condition that will determine $c$ must be imposed in the asymptotic region. This is a familiar strategy in holography: any sensible condition must be imposed near the boundary and not in the deep IR region of the geometry, where the horizon is located. Actually, in asymptotically AdS spaces (and contrary to what happens in asymptotically flat spaces), we also expect that the condition we need to impose will still allow some mild ambiguity in $\Sb^{*}$. Indeed, in view of the fundamental relation \eqref{fundid} we wish to prove, we should be allowed to add arbitrary finite local counterterms. In the present case, the most general counterterm action, constrained by locality, general covariance and power counting, is of the form
\be\label{SCTex} \SCT = \beta a^{3}\Bigl(\frac{c_{0}}{L^{4}} + \frac{c_{1}}{L^{2}a^{2}} + \frac{c_{2}}{a^{4}}\Bigr)\, ,\ee
for dimensionless renormalization constants $c_{0}$, $c_{1}$ and $c_{2}$ that may depend on a regulator $\epsilon$ but not on $a$ or $\beta$. These terms correspond to adding a cosmological constant, curvature and curvature squared terms in the boundary theory. 

These considerations yield the following simple proposal to fix the ambiguity associated with the integration constant $s$:

\noindent\emph{The brane action, evaluated for a brane worldvolume $r=\epsilon$, where $r$ is the Fefferman-Graham radial coordinate and $\epsilon>0$ a regulator, should go to a purely counterterm action near the boundary, up to terms that go to zero when $\epsilon\rightarrow 0$.}

\noindent This is a very natural prescription and we believe that it is the only consistent one. Moreover, it is compatible with the construction in \cite{fer1} and is in harmony with the general intuition that going to the boundary of bulk space corresponds to a UV limit in the boundary field theory. In our example, using the well-known formulas for the tension of a D3-brane in type IIB string theory and the relation between the AdS scale $L$ and the string theory parameters \cite{malda},
\be\label{stanrel} \tau_{3}= \frac{1}{2\pi \ell_{\text s}^{4}g_{\text s}}\, \cvp\quad L^{4}= \frac{\ell_{\text s}^{4}g_{\text s} N}{\pi}\,\cvp\ee
we see that the first term in the right-hand side of \eqref{Sbex} is precisely of the form \eqref{SCTex} when $x\rightarrow 0$. The constant $s$ must thus be of the form \eqref{SCTex} as well. Putting everything together, we obtain
\be\label{Sbexfinal} \Sb^{*} = \frac{N\beta}{8 a}\, \frac{4\alpha -1}{\alpha^{2}} + \SCT\, ,\ee
for an arbitrary finite counterterm action $\SCT$. Using \eqref{fundid}, with an exponent $\gamma=2$ suitable for a free energy scaling as $N^{2}$ in gauge theory, we reproduce precisely the correct free energy of the $\nn=4$ Yang-Mills theory \cite{shol2,WSch}.

\noindent\emph{Remark}: the shrinking of the probe brane to the tip of the cigar geometry might be interpreted as the Euclidean version of a brane falling into the horizon of the Minkowskian black hole geometry. However, we would like to emphasize that this is misleading. As will be clear in the next section, the tip of the cigar is not a special point for the brane. If allowed to deform in arbitrary ways, the brane can shrink at any point on the cigar, thus including at $r<\rh$. Only the minimal value of $\Sb$ has a physical meaning.

\section{\label{GenSec} The general case}

Let us now consider an arbitrary Poincar\'e-Einstein bulk space $M$. We pick a representative $\bar g$ of the conformal structure on the boundary $X=\partial M$. We denote by $r$ the Fefferman-Graham radial coordinate and by $z$ the coordinates on $X$. The bulk metric near the boundary reads
\be\label{PEmet} G = \frac{L^{2}\d r^{2}+ g}{r^{2}}\, \cvp\ee
where
\be\label{FGexp} g(r,z)=\bar g(z) + g_{(2)}(z) r^{2} + \cdots\ee
has the usual near-boundary Fefferman-Graham expansion. We introduce a regulator $\epsilon>0$, denote by $\Sigma_{\epsilon}$ the hypersurface $r=\epsilon$ and by $M_{\epsilon}$ the interior of $\Sigma_{\epsilon}$, the regulated bulk space. We also denote with the symbol $\equiv$ equalities modulo the addition of local counterterms on the boundary and terms that go to zero when $\epsilon\rightarrow 0$. The gravitational action is the sum of the Einstein-Hilbert and the Gibbons-Hawking terms, which is a surface integral over $\Sigma_{\epsilon}$. It is easy to check, using the expansion \eqref{FGexp}, that the Gibbons-Hawking term is always a pure counterterm. This is a nice consequence of using the Fefferman-Graham coordinate $r$ to regulate the bulk space. Using Einstein's equations \eqref{PEcond}, we thus obtain
\be\label{Sgform} \Sg^{*} \equiv -\frac{1}{16\pi G_{d+1}}\int_{M_{\epsilon}}\!\d^{d+1}x\sqrt{\det G}\,\Bigl(R +\frac{d(d-1)}{L^{2}}\Bigr) = \frac{d}{8\pi G_{d+1}L^{2}} V(M_{\epsilon})\, ,\ee
where $G_{d+1}$ is the bulk Newton constant and $V(M_{\epsilon})$ the volume of the regulated bulk space.

We now have to define what we mean by probe brane in general. On physical grounds, it is reasonable to consider that a probe brane should be an embedding of the boundary manifold $X$ in $M_{\epsilon}$ which can be obtained by smoothly deforming $\Sigma_{\epsilon}$. A less stringent requirement would be to consider all hypersurfaces homologous to the boundary. We shall work with this second point of view for simplicity, but we believe that the first point of view should be equivalent for our purposes (at least it is on the specific examples we are aware of). We denote by $M_{\Sigma}$ the bulk space enclosed by $\Sigma$, $\partial M_{\Sigma} = \Sigma$. The DBI term in the brane action is simply $\tau_{d-1} A(\Sigma)$, where $\tau_{d-1}$ is the brane tension and A its area (worldvolume) for the induced metric on $\Sigma$. The Euclidean CS term is $-i\tau_{d-1}\int_{\Sigma}\! C_{d}$, with $\d C_{d} = \frac{id}{L}\Omega_{d+1}$ proportional to the volume form of the bulk space, generalizing \eqref{SCSdef} and \eqref{RR5form}. Integrating to get $C_{d}$ produces an arbitrary integration constant $s$, as in the example of section \ref{ExSec}. This constant must be the same for all the probe branes, since they are all homologous to each other. Moreover, up to this constant, Stokes' theorem implies that the CS term is proportional to the volume of $M_{\Sigma}$. Overall, we thus obtain
\be\label{Sbgene} \Sb (\Sigma) = \tau_{d-1}\Bigl(A(\Sigma) - \frac{d}{L}V(M_{\Sigma})\Bigr) + s\, .\ee
The constant $s$ is fixed by using the principle formulated in Sec.\ \ref{ExSec}: we impose that $\Sb(\Sigma_{\epsilon})\equiv 0$, i.e.\ the brane action on the boundary is a pure local counterterm. Since it is obvious that $A(\Sigma_{\epsilon})\equiv 0$, the area being a local cosmological constant term on the boundary, we get, by taking \eqref{Sgform} into account,
\be\label{Sbgene2} \Sb (\Sigma) \equiv \tau_{d-1}\Bigl(A(\Sigma) - \frac{d}{L}V(M_{\Sigma})\Bigr) + 8\pi G_{d+1}L\tau_{d-1}\Sg^{*}\, .\ee
To compute the on-shell brane action $\Sb^{*}$, \emph{we thus have to minimize the functional $A - \frac{d}{L}V$ over all probe branes.} If we can prove the isoperimetric inequality \eqref{isop}, then the minimum value will be zero, which is realized by a shrunken brane.\footnote{In the explicit examples we know, it is clear that the boundary can always be shrunk to zero area. More generally, this follows easily if one knows the topology of the bulk, as in theorems by Graham and Lee and Lee \cite{GLL}. Even more generally, this follows from the fact that the boundary $\Sigma_{\epsilon}=\partial M_{\epsilon}$ is in a trivial homology class.} The identity \eqref{fundid} would automatically follow, with an exponent
\be\label{gammaform} \gamma = 8\pi L G_{d+1}\tau_{d-1}N\, .\ee
Note that $\gamma$ must be independent of $N$. For example, in type IIB with the $\nn=4$ theory living on the boundary $X$, the ten dimensional Newton constant is $G_{10}=\smash{\frac{1}{2}\pi^{2}\ls^{8}\gs^{2}}=\smash{\frac{\pi^{4}L^{8}}{2N^{2}}}$, and thus, taking into account the volume $\pi^{3}L^{5}$ of the $\text{S}^{5}$ piece in the geometry, the five dimensional $G_{5}=\smash{\frac{\pi L^{3}}{2N^{2}}}$. Using \eqref{stanrel}, we see that \eqref{gammaform} yields $\gamma=2$ as expected. Cases with other values of $\gamma$ are discussed in \cite{FR}. 

Thus there remains to understand the crucial inequality \eqref{isop}. This kind of inequalities have been much studied in mathematics, see e.g.\ \cite{Chavel}. The infimum of the ratios $A(\Sigma)/V(M_{\Sigma})$ is known as the Cheeger constant $I_{\infty}(M)$ of the non-compact manifold $M$. The inequality \eqref{isop} is thus equivalent to a lower bound for the Cheeger constant, $I_{\infty}(M)\geq d/L$. Interestingly, it is known that $I_{\infty}^{2}/4$ provides a lower bound on the spectrum of the Laplacian on $M$ \cite{Chavel}. The inequality \eqref{isop} thus also implies a generalized Breitenlohner-Freedman bound.

To build intuition on \eqref{isop}, it is very instructive to start by considering a special class of large hypersurfaces. We use the Trudinger-Aubin-Schoen theorem to pick a conformal class representative $\bar g$ on the boundary having constant scalar curvature $\bar R$. It is then straightforward to compute $A - \frac{d}{L}V$ for hypersurfaces $\Sigma$ given by $r=\text{constant}$, at small $r$, where $r$ is the Fefferman-Graham radial coordinate associated with $\bar g$, by using the expansion \eqref{FGexp}. One finds that it diverges as $\bar R/r^{d-2}$, if $d>2$, or as $-\bar R\ln r$ if $d=2$ \cite{WY, Yampos}. In particular, if $\bar R<0$, the probe brane action is unbounded from below and $\Sb^{*}=-\infty$! A crucial requirement is thus that $\bar R\geq 0$. This is equivalent to saying that the Yamabe constant $Y([\bar g])$ of the conformal class at infinity is non-negative. Holography will be inconsistent in such cases, precisely due to the emission of large probe branes, as argued in \cite{Yampos,Rabadan}.\footnote{See \cite{unstableref} and references therein for interesting recent physical applications of this instability.} Let us emphasize that this argument shows that the stability of the bulk string theory implies that $A(\Sigma) - \frac{d}{L}V(M_{\Sigma})$ must be bounded from below. For our purposes, we need the much stronger result \eqref{isop} to be true: that the bound should always be strictly zero.

We thus limit ourselves to the cases $Y[\bar g]\geq 0$. Remarkably, the inequality \eqref{isop} was then derived in \cite{Wang}, building on the results in \cite{WY} and on geometric measure theory. Let us sketch here a more elementary approach, based on some of the results of \cite{Lee}. The idea is to consider a scalar field $\phi$ on $M$ of mass $m^{2}= (d+1)/L^{2}$, thus sourcing an operator of dimension $\delta = d+1$ on the boundary. As usual, such a scalar field will behave as $r^{d-\delta}=1/r$ near the boundary. For our purposes, we choose the source $\bar\phi =\lim_{r\rightarrow 0}(r\phi)$ to be a strictly positive constant, say equal to one. The field equation $(\Delta + m^{2})\phi=0$ then implies immediately, from the maximum principle, that $\phi>0$ on $M$. Moreover, using \eqref{PEcond}, it is not difficult to check that $\Delta(|\d\phi|^{2}-\phi^{2}/L^{2})\leq 0$, where $|\d\phi|^{2}=G^{\mu\nu}\partial_{\mu}\phi\partial_{\nu}\phi$. The maximum principle then implies that
\be\label{fundiphi} |\d\phi|^{2}- \phi^{2}/L^{2}\leq 0\ee
on $M$, as soon as this is valid near the boundary $r=0$. But, when $r\rightarrow 0$, this inequality can be directly checked by using the expansion \eqref{FGexp} and the similar well-known expansion for the scalar field. Using the same conformal class representative as in the previous paragraph, with constant scalar curvature $\bar R$, one finds that $|\d\phi|^{2}- \phi^{2}/L^{2}\simeq -\bar R/(d(d-1))$ near the boundary, which is indeed non-positive if $Y([\bar g])\geq 0$.

This being established, we can proceed as follows.\footnote{We are grateful to Gilles Carron for providing this simple argument.} We consider the vector field $v^{\mu}=\partial^{\mu}\ln\phi$. By using \eqref{fundiphi}, we immediately find that
\be\label{inev} |v|^{2}=G_{\mu\nu}v^{\mu}v^{\nu}\leq \frac{1}{L^{2}}\, \cvp\quad
\div v = \nabla_{\mu}v^{\mu}\geq\frac{d}{L^{2}}\,\cdotp\ee
We now integrate the second inequality above over $M_{\Sigma}$ and use Stokes' theorem to find
\be\label{inev2} \int_{\Sigma}\d^{d} x\sqrt{\text{P}(G)}\, v^{\mu}n_{\mu}\geq \frac{d}{L^{2}}V(M_{\Sigma})\, ,\ee
where $\text{P}(G)$ denotes the determinant of the induced metric on $\Sigma$ and $n$ is the unit normal to $\Sigma$, pointing outward. The isoperimetric inequality \eqref{isop} then follows from the bounds
\be\label{bounds}0\leq \int_{\Sigma}\!\d^{d} x\sqrt{\text{P}(G)}\, v^{\mu}n_{\mu}\leq \int_{\Sigma}\!\d^{d} x\sqrt{\text{P}(G)}\, \bigl|v^{\mu}n_{\mu}\bigr|\leq \int_{\Sigma}\!\d^{d} x\sqrt{\text{P}(G)}\, |v|\leq \frac{A(\Sigma)}{L}\,\cvp\ee
where, in the last step, we have used the first inequality in \eqref{inev}.
 

%
\subsection*{Acknowledgments}

We would like to thank C.~Bachas, G.~Carron, J.~Fine, E.~Kiritsis and I.~Sachs for very useful discussions. The work of F.F.\ is supported in part by the Belgian FRFC (grant 2.4655.07) and IISN (grants 4.4511.06 and 4.4514.08). The work of A.R.\ is supported in part by the DFG Transregional Collaborative Research Centre TRR 33 and the DFG cluster of excellence ``Origin and Structure of the Universe'' as well as the Belgian American Educational Foundation.


\begin{thebibliography}{99}
%

%
\bibitem{malda}{J.~Maldacena, \atmp{2}{1998}{231}, hep-th/9711200.}
%
\bibitem{KT}{E.~Kiritsis and T.R.~Taylor, \emph{PoS trieste99} (1999) 027, hep-th/9906048,\\
E.~Kiritsis, \jhep{10}{1999}{010}, hep-th/9906206,\\
A.A.~Tseytlin and S.~Yankielowicz, \npb{541}{1999}{145}, hep-th/9809032,\\
S.~Ryang, \mpla{14}{1999}{1573}, hep-th/9903157.}
%
\bibitem{fer1}{F.~Ferrari, \npb{880}{2014}{247}, arXiv:1310.6788,\\
F.~Ferrari, \npb{880}{2014}{290}, arXiv:1311.4520,\\
F.~Ferrari, \prd{89}{2014}{105018}, arXiv:1308.6802.}
%
\bibitem{fer2}{F.~Ferrari, \npb{869}{2013}{31}, arXiv:1207.0886,\\
F.~Ferrari, \npb{871}{2013}{181}, arXiv:1301.3722,\\
F.~Ferrari, M.~Moskovic and A.~Rovai, \npb{872}{2013}{184}, arXiv:1301.3738,\\
F.~Ferrari and M.~Moskovic, \plb{723}{2013}{455}, arXiv:1301.7062,\\
F.~Ferrari and A.~Rovai, \plb{724}{2013}{121}, arXiv:1303.7254,
E.~Conde and M.~Moskovic, \jhep{04}{2014}{148}, arXiv:1312.0621.}
%
\bibitem{HEA}{J.H.~Schwarz, \jhep{01}{2014}{088}, arXiv:1311.0305.}
%
\bibitem{shol1}{S.~Gubser, I.R.~Klebanov and A.M.~Polyakov, \plb{428}{1998}{105}, hep-th/9802109.}
%
\bibitem{shol2}{E.~Witten, \atmp{2}{1998}{253}, hep-th/9802150.}
%
\bibitem{FR}{F.~Ferrari and A.~Rovai, \emph{On-Shell D-Brane Action and Holography}, to appear.}
%
\bibitem{holoren}{V.~Balasubramanian and P.~Kraus, \cmp{208}{1999}{413}, hep-th/9902121,\\
R.~Emparan, C.V.~Johnson and R.C.~Myers, \prd{60}{1999}{104001}, hep-th/9903238,\\
C.R.~Graham, Proc.\ of the 19th Winter School in Geometry and Physics, Srni, Czech Rep., 1999, math/9909042,\\
S.\ de Haro, K.~Skenderis and S.N.~Solodukhin, \cmp{217}{2001}{595}, hep-th/0002230,\\
K.~Skenderis, \cqg{19}{2002}{5849}, hep-th/0209067.}
%
\bibitem{WY}{E.~Witten and S.-T.~Yau, \atmp{3}{1999}{1635}, hep-th/9910245.}
%
\bibitem{Yampos}{J.~Maldacena, J.~Michelson and A.~Strominger, \jhep{02}{1999}{011}, hep-th/9812073,\\
N.~Seiberg and E.~Witten, \jhep{04}{1999}{017}, hep-th/9903224.}
%
\bibitem{Lee}{J.M.~Lee, \cag{3}{1995}{253}, dg-ga/9409003.}
%
\bibitem{Wang}{X.~Wang, \cag{10}{2002}{647}.}
%
\bibitem{WSch}{E.~Witten, \atmp{2}{1998}{505}, hep-th/9803131.}
%
\bibitem{GLL}{C.R.~Graham and J.M.~Lee, \amath{87}{1991}{186},\\
J.M.~Lee, \emph{Mem.\ Amer.\ Math.\ Soc.\ }\textbf{183} (2006), math.DG/0105046.}
%
\bibitem{Rabadan}{M.~Kleban, M.~Porrati and R.~Rabadan, \jhep{08}{2005}{016}, hep-th/0409242.}
%
\bibitem{Chavel}{I.~Chavel, \emph{Riemannian geometry -- A modern introduction,} Cambridge U.\ Press, 1993,\\
T.~Sakai, \emph{Riemannian Geometry,} Translations of Mathematical Monographs, vol.\ 149, Shokabo Publishing Co., 1992, Am.\ Math.\ Soc., 1996.}
%
\bibitem{unstableref}{B.~McInnes, \cqg{31}{2014}{025009}, arXiv:1211.6835,\\
Y.~C.~Ong and P.~Chen, \jhep{07}{2013}{147}, arXiv:1304.3803,\\
B.~McInnes and E.~Teo, \npb{878}{2014}{186}, arXiv:1309.2054,\\
Y.~C.~Ong, B.~McInnes and P.~Chen, arXiv:1403.4886,\\
B.~McInnes, arXiv:1409.3663.}
%
\end{thebibliography}
\end{document}